\documentclass{article}
\usepackage{spconf,amsmath,graphicx}
\usepackage{multirow}

\title{Video Logo Retrieval based on local Features}
\name{Bochen~Guan$^{\star}$ \qquad Hanrong~Ye$^{\dagger}$ \qquad Hong Liu$^{\dagger}$ \qquad William A. Sethares$^{\star}$
\thanks{Bochen Guan and Hanrong Ye contributed equally to this work.  Hanrong Ye and Hong Liu are supported by National Natural Science Foundation of China (NSFC, No. U1613209).}}

\address{$^{\star}$Department of Electrical and Computer Engineering, University of Wisconsin-Madison, WI\\
$^{\dagger}$Key Laboratory of Machine Perception, Peking University, China.}
\def\alg{VLR}

\begin{document}
%
\maketitle
\begin{abstract}
Estimation of the frequency and duration of logos in videos is important and challenging in the advertisement industry as a way of estimating the impact of ad purchases. Since logos occupy only a small area in the videos, the popular methods of image retrieval could fail. This paper develops an algorithm called Video Logo Retrieval (\alg), which is an image-to-video retrieval algorithm based on the spatial distribution of local image descriptors that measure the distance between the query image (the logo) and a collection of video images. \alg~uses local features to overcome the weakness of global feature based models such as convolutional neural networks (CNN). Meanwhile, \alg~is flexible and does not require training after setting some hyper-parameters. The performance of \alg~is evaluated on two challenging open benchmark tasks (SoccerNet and Standford I2V), and compared with other state-of-the-art logo retrieval or detection algorithms. Overall, \alg~shows significantly higher accuracy compared with the existing methods.
\end{abstract}
\section{Introduction}
\label{sec:intro}

Algorithms such as \cite{chen2018gated, 1631272, revaud2012correlation, liao2017mutual, rootsift, deeplocal1, delf, liu2019fully} that can retrieve, detect, or localize a target logo in a video stream. \cite{revaud2012correlation} have applications in automatic annotation, dissemination impact evaluation, and suspicious logo detection  \cite{1631272, zhengliang}. Though several methods \cite{chen2018gated, deeplogo} show good performance for logo retrieval, their accuracy in broadcast videos are not acceptable for large scale commercial adoption \cite{1631272, revaud2012correlation}. Logos typically occupy only a small proportion of the frame, which can lead to failure of popular global signature-based methods such as SIFT+VLAD \cite{vlad}, SIFT+FV \cite{fv} and CNNs \cite{zhengliang}. Techniques for local descriptor matching can locate locally similar images, but suffer from a high false positive rate when used with highly semantic video frames \cite{liao2017mutual}. Logo targets are often updated frequently and so there may be inadequate training data for the application of logo detection algorithms. Many image retrieval studies are well adapted for image--to--image searching, but lack systematic estimation for image--to--video retrieval.



\begin{figure*}[htbp]
\centering
\includegraphics[scale=0.2]{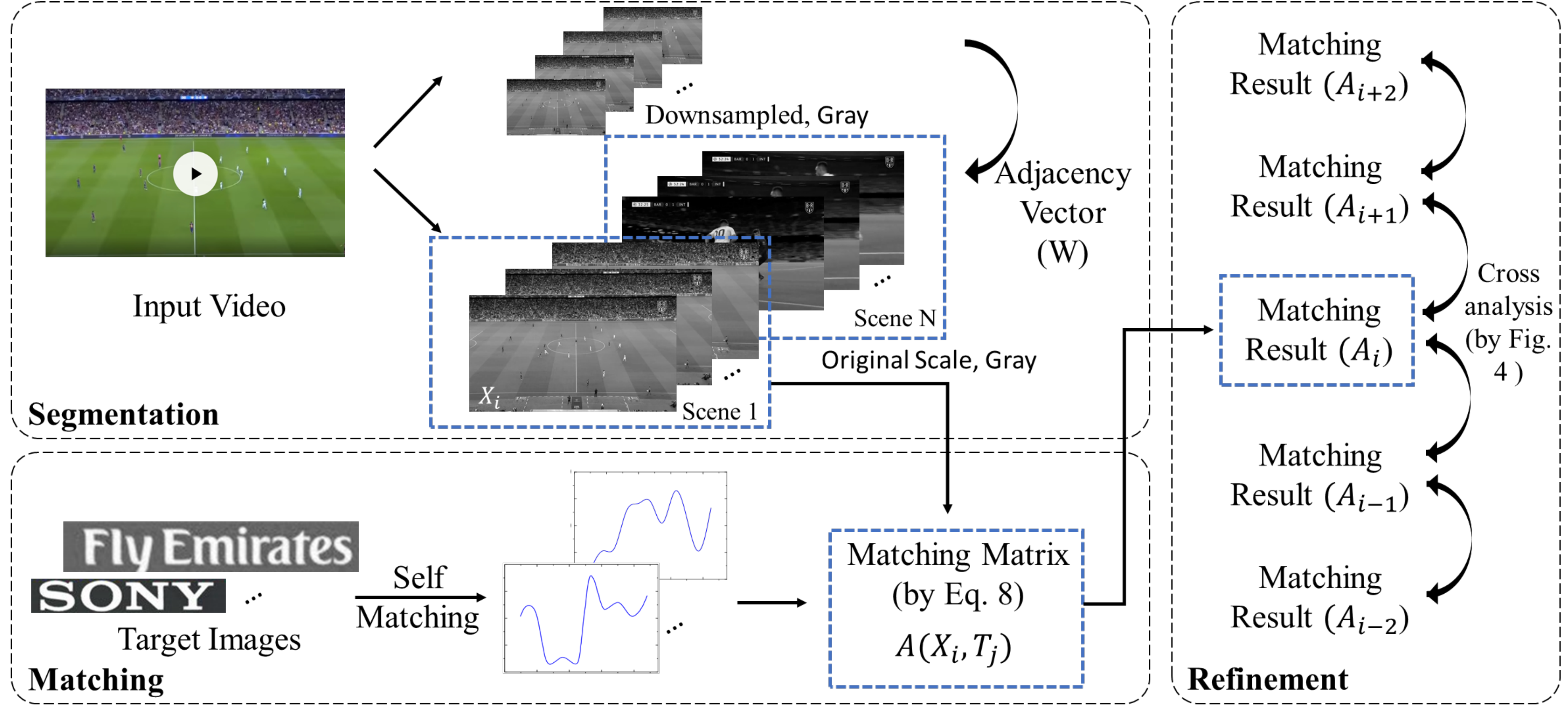}
\caption{The \alg~algorithm consists of three stages: video segmentation, matching, and refinement. Local features are transformed by using \eqref{eqn:matchProb} to generate an adjacency vector that splits the original video into scenes. Eq. \eqref{eq:targetp}-\eqref{chunk} are used to match target logos with videos. The refinement stage smooths the results and maintains continuity.}
\label{Fig1}
\end{figure*}

This paper proposes the \alg~algorithm for Logo Retrieval outlined in Fig. \ref{Fig1}. \alg~contains three stages: segmentation, matching, and refinement. The process begins by down-sampling a video stream and segmenting it into regions that can be interpreted as corresponding to different camera angles or different scenes where objects can be assumed to move continuously. The matching stage analyzes individual frames (full scale) and compares to each of the target images using local feature descriptor matching, resulting in the matching matrix. 
The refinement stage then performs a cross analysis of all video images in the scene and exploits the continuity of information over time. 

\section{Related Work} 

Content based image retrieval usually includes both hand-crafted descriptor approaches and deep learning approaches \cite{zhengliang}. Before 2012, Bag-of-Words (BoW) models with hand-crafted features were predominant, including descriptors such as SIFT, SURF and their variants \cite{sift, surf, pcasift, rootsift}. These BoW models compute statistics of the descriptors and encode images in a compact feature vector. Other BoW-like methods include VLAD \cite{vlad} and FisherVector \cite{fv}, which show better performance in some applications. 


With the rapid development of convolutional neural networks\cite{chen2018gated}, deep learning approaches have come to dominate \cite{delf, liao2017mutual}. 
The performance of the deep learning methods has been improved with the appearance of strong backbone models such as Resnet \cite{resnet}. 
One promising method is DELF \cite{delf}, which designs an attention layer to extract and select the semantic local features from images, and which shows good performance compared with hand crafted descriptors. Recently, there are some interesting works focusing on logo detection in videos \cite{liao2017mutual,deeplogo}. Unlike retrieval algorithms, these methods adopt deep features in an object detection fashion. 


\section{Video Logo Retrieval via \alg} 
\label{sec:description}


The \alg~algorithm consists of the three stages outlined in Fig. \ref{Fig1}. Suppose the input is a source video with $n$ frames, and that there are $m$ target images of dimension $(H_t, W_t)$. The goal is to locate these target images in the input video. 

\subsection{Segmentation} 

Videos typically contain many scenes that have little correlation. The video segmentation stage of \alg~is used to segment the video into individual scenes that can be analyzed separately. We adopt a simple but effective method for video segmentation using local feature descriptors, although other scene change detection algorithms \cite{scene_change_motion} may also be applied. 


The video segmentation stage contains the two steps shown in Fig. \ref{Fig1}. First, to increase the computational efficiency, each frame is converted to grayscale and down-sampled. Let $\bar{X}_{i}$ be the down-sampled $i$th frame. Next, compute local features and the matches between consecutive pairs: $\bar{X}_{i}$ and $\bar{X}_{i+1}$. 
This measure of similarity can be formalized by counting the number of detected local feature matching points (matched local feature points between two consecutive frames) divided by the total number of local feature matching keypoints.
Accordingly, let 
\begin{equation} 
\label{eqn:matchProb}
p(\bar{X}_{i},\bar{X}_{i+1})=\frac{\text{\# of matching points}}{\text{total \# of keypoints}}.
\end{equation}
Each term \eqref{eqn:matchProb} lies between 0 and 1, and can be interpreted as a probability of two successive frames belonging to the same video scene. These are concatenated into the adjacency vector:
\begin{align}
W    &= \left( p(\bar{X}_{1},\bar{X}_{2}), p(\bar{X}_{2},\bar{X}_{3}) ,\ldots,  p(\bar{X}_{n-1},\bar{X}_{n}) \right) \nonumber \\
     &= \left( w_1,w_2,...,w_{n-1} \right),
\label{label vector}
\end{align}
which represents the successive similarities over time. Each element of $W$ computes the match between neighboring frames. Fig. \ref{Fig2} shows an example, where (a) is the matching between two frames and (b) is the probability distribution between those neighboring frames. If a scene change or camera cut occurs, the number of matching keypoints will decrease. 
The Page-Lorden CUSUM algorithm \cite{hawkins2014cusum} is used to detect change points in the video and to split the video frames into a collection of scenes. The following stages analyze each scene separately. 

\subsection{Matching}
\label{sec:typestyle}
 
\begin{figure}[htbp]
    \centering
    \includegraphics[scale=0.23]{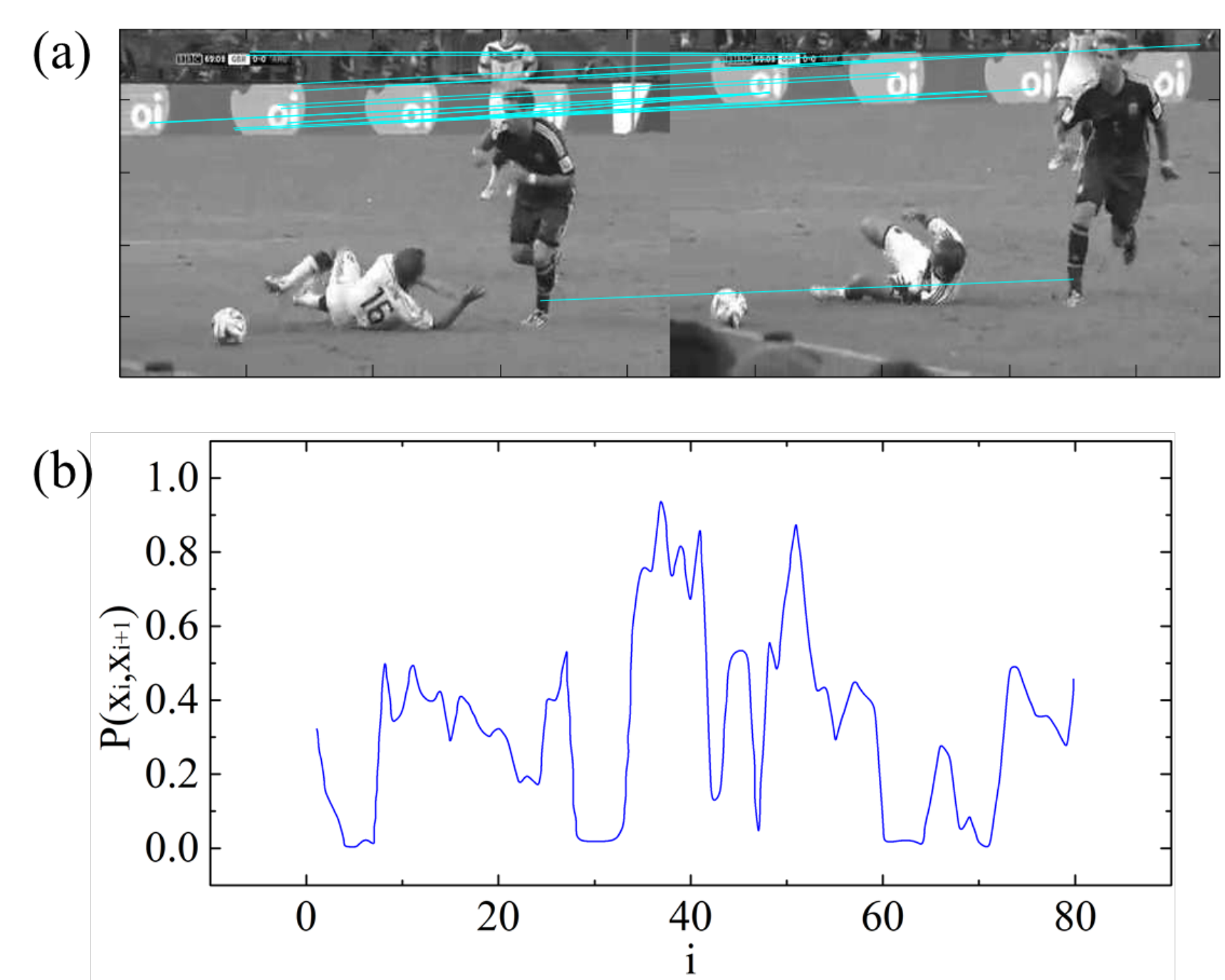}
    \caption{(a) shows the matching points between two neighboring images. (b) shows the matching score as a function of time. The images are divided into different segments by using Page-Lorden CUSUM algorithm \cite{hawkins2014cusum}}
    \label{Fig2}
\end{figure}

\begin{figure}[htbp]
    \centering
    \includegraphics[scale=0.1]{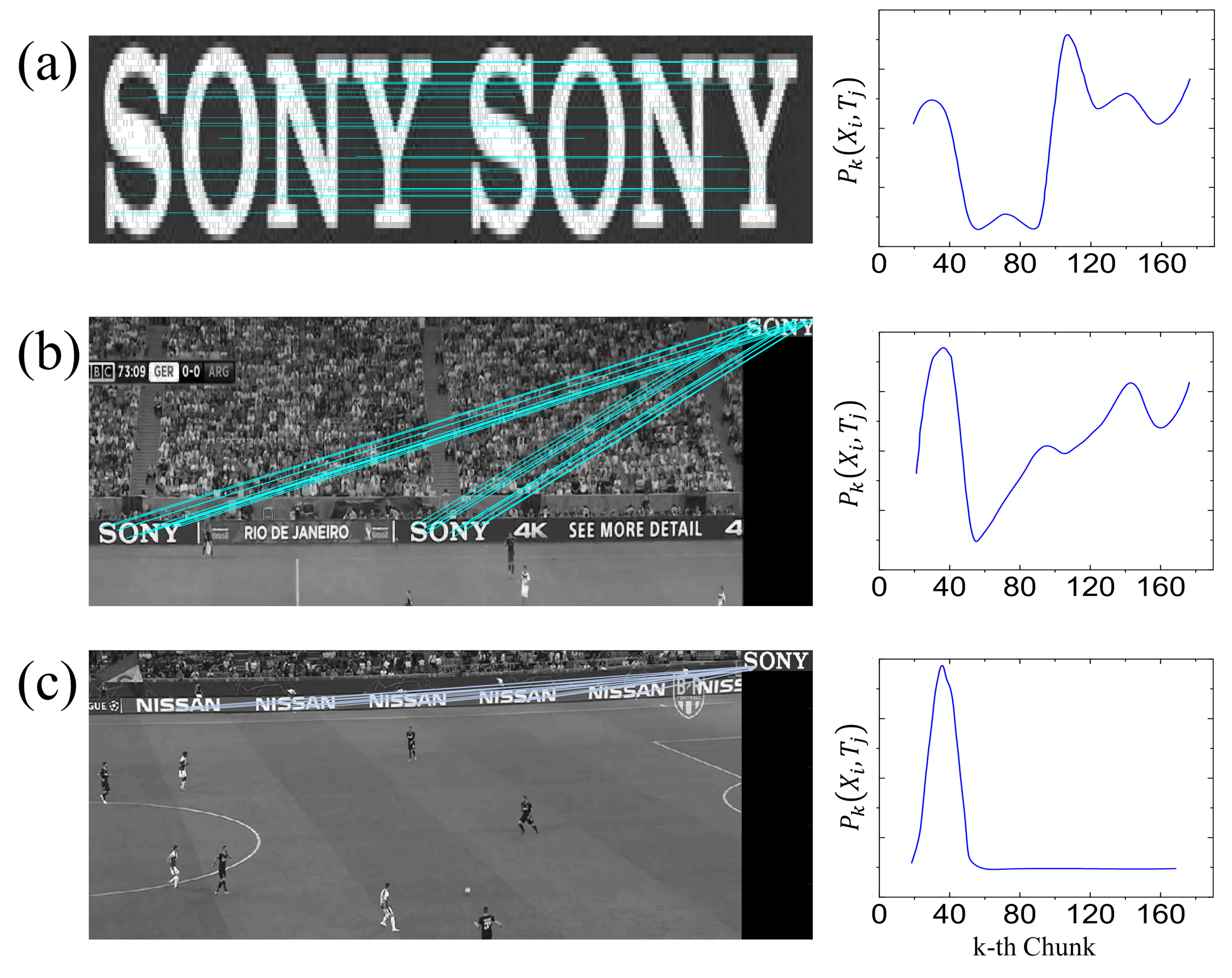}
    \caption{(a) shows a matching of a target image with itself. (b) shows a matching between a target image and a video frame. (c) shows an incorrect matching between a target image and a video frame despite many matching points (all the matching points are concentrated).}
\label{Fig3}
\end{figure}

The matching stage uses SIFT or other comparable local feature sets to compute a matching matrix. Some examples of matches between targets and video images are shown in Fig. \ref{Fig3}. Intuitively, targets with the most matching points will be desirable candidates. Let the number of keypoints in the target $j$ be $N(T_j)$ and the number of matching points between $T_j$ and the source video image $i$ (i.e., $X_i$) be $N(T_j, X_i)$. As in Eq. \eqref{eqn:matchProb}, the empirical probability of $T_j$ given $X_i$ is
\begin{equation}
  p(X_i, T_j) = \frac{N(T_j, X_i)} {N(T_j)} .
\label{eq:targetp}
\end{equation}
In Fig.~\ref{Fig2}(c), though $p(X_i, T_j)$ is high, it is obviously not a good matching result. Additional factors are needed. 
 
Suppose the image sequence within a given scene has $h$ images: $X_{1}, X_{2}, \ldots, X_{h}$. Partition each target image into $P$ vertical chunks: $s = 1,..., P$. If the target is of size $L \times K$, \alg~divides the target images into $P$ chunks of size $\frac{L \times K}{P}$.
We establish a set of ground values by matching SIFT keypoints between the targets with themselves, which is shown in Fig. \ref{Fig2}. Let $M_k(T_j)$ be the number of self-matching points in chunk $k$. The self-matching probability for $T_j$ in the $k$th chunk is
\begin{equation}
p_k(T_j,T_j)=\frac{M_k(T_j)}{N(T_j,X_i)}.
\label{chunk}
\end{equation}
Since each keypoint is also a matching point for the self-matching situation,
\begin{equation}
p(T_j,T_j) = \sum_{k} p_k(X_i,T_j)=1,
\end{equation}
which suggests that $p(T_j,T_j)$ can be interpreted as a matching probability distribution. This can be used as a reference when comparing the target images to other video source images.

The matching between the individual video images and the target images is again conducted using (\ref{chunk}). So we can similarly use $p_k(X_i,T_j)$ to describe the keypoints matching between the target images (with $P$ chunks) and the video images. Since not all keypoints can be matched,
\begin{equation}
p(x_i,T_j) = \sum_{k} p_k(x_i,T_j) \leq 1,
\end{equation}
where $p(x_i,T_j)$ is the probability of $T_j$ with $X_i$. Comparing with the self-matching distribution, the correlation between these two distributions will determine if the video image contains the target. More precisely, if the correlation is weak, the target image $T_j$ is unlikely to be in the image frame $X_i$ even if there are some matching points. The Kullback--Leibler (KL) divergence of the distributions is used to describe the correlation
\begin{equation}
D_{KL}(X_i,T_j)=\sum_{k} p_k(X_i,T_j)\log{\frac{p_k(X_i,T_j)}{p_k(T_j,T_j)}}.
\end{equation}
Define the matching matrix $\bf A$ between the video image $X_{i}$ and the target image $T_{j}$ as
\begin{equation} 
{\bf A}_{i,j}= [p(X_i,T_j), D_{KL}(X_i,T_j)] .
\label{matrix}
\end{equation}
Then \alg~uses $k$-$means$ clustering to pick highly distinctive matching scores from matching matrix $A$. If the score of both $p(X_i,T_j)$ and $D_{KL}(X_i,T_j)$ are high, the image is said to contain the target image.

\subsection{Refinement} \label{sec:refinement}

\begin{figure}[htbp]
    \centering
    \includegraphics[scale = 0.1145]{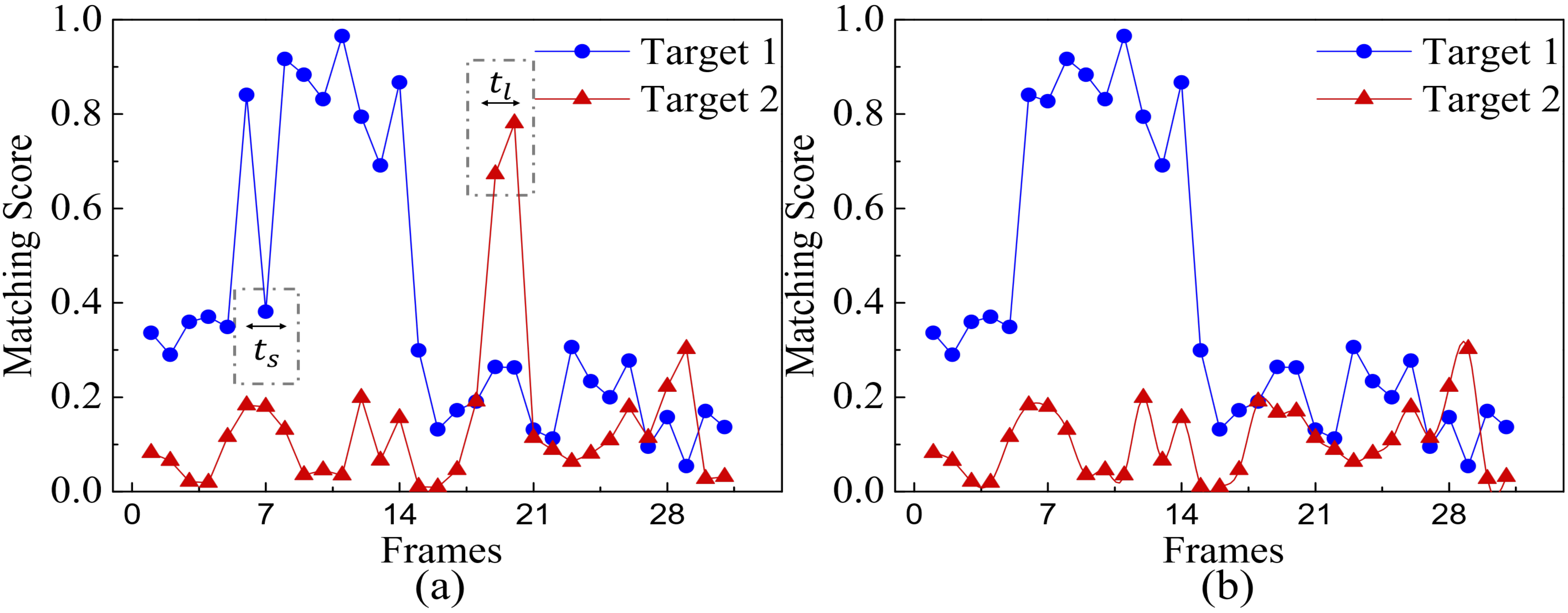}
    \caption{A sketch of two conditions in the refinement stage. (a) is the matching result for two different targets in a list of video frames. The time interval of targets has a sudden change. The cross analysis refines the matching result in (b).}
    \label{Estimation_stage}
\end{figure}

The refinement stage conducts a cross analysis of adjacent video images to determine which target images are present. Within a single continuous scene, it is expected that targets will generally persist for a significant number of frames; it is unreasonable for a target to appear, disappear, and re-appear in a short time. Thus the raw matching result should be smoothed by this assumed continuity over time. Suppose that a sequence of video images $X_i$ lie in a scene. We develop two parameters to smooth the estimation scores and improve the matching result: the minimum stand time $t_{s}$ and the maximum lost time $t_{l}$. This cross analysis process is illustrated in Fig. \ref{Estimation_stage}.

\noindent
\textbf{The Minimum Standing Time $t_{s}$}. If a target image appears as a candidate match for a time less than $t_{s}$, it is rejected. On the other hand, if the target image persists for a time greater than $t_{s}$, it is accepted.

\noindent
\textbf{The Maximum Lost Time $t_{l}$}.  Suppose a candidate match $T_j$ is present but absent for time from $i$ to $i+t$, If the interval $t$ is less than $t_{l}$, then it is presumed that this absence is due to noise or other transient issues and it should be considered present for the complete time.


\section{Experiments}
\label{Experiments}
This section demonstrates the feasibility of  \alg~using two benchmark datasets. We test \alg~on two different local feature extractors (SIFT, DELF) and compare the performance with several state-of-the-art algorithms: a CNN based global features algorithm (ResNet-101) \cite{resnet}, SIFT\cite{sift}, DELF\cite{delf}, logo recognition algorithm\cite{deeplogo}, and a BoW-like algorithm (VLAD) \cite{vlad}.

\subsection{Dataset and implementation details}

\textbf{SoccerNet} is a dataset of soccer games of European Championships from 2014 to 2017\cite{giancola2018soccernet}. We randomly selects 45 UEFA games and down-sample one frame a second from the source video images, testing a total of 248,400 images with resolution of 1920$\times$1080 pixels. Since the sponsors of UEFA vary for different years, we use seven advertisements sponsored each year from 2014 to 2017 as the targets. 

\textbf{Stanford I2V} is a video database consisting of newscast videos annotated with more than 200 ground-truth queries \cite{Araujo1}. Our experiment selects 200 news videos, most last no longer than 2 minutes. All the videos are down-sampled to one frame per half second. 


\alg~was implemented using SIFT features in Matlab 2016a. For \alg~of DELF, features are first computed using Python/Tensorflow, and the matching is computed by using the same technique as in Section \ref{sec:description}. The comparison of experiments of VLAD and SIFT are conducted in Matlab, while the deep learning algorithms are run in Python/Tensorflow. Target logos are usually updated every season/year. Since there are only a limited images of new logos, they cannot provide an adequate dataset for training alone. Instead, the deep learning models are trained using the OpenLogo dataset \cite{agarwal2018recent}, which includes many of the target logo images needed in the experiments. 

\subsection{Results}
\begin{table}
\centering
\caption{Comparison of proposed techniques against state-of-the-art algorithms on SoccerNet dataset in three different seasons.}

\label{table:UEFA}
\scalebox{0.83}{
\begin{tabular}{c|c|c|c} 
\hline
\multirow{2}{*}{Model}    & \multicolumn{3}{c}{SoccerNet (mAP)}  \\ 
\cline{2-4}
                          & 2014-2015 & 2015-2016 & 2016-2017    \\ 
\hline
ResNet101-GeM~\cite{resnet}            & 67.5      & 65.2      & 73.6         \\
SIFT~\cite{sift}                      & 69.5      & 65.3      & 75.2         \\
VLAD \cite{vlad}                      & 70.6      & 66.4      & 72.8         \\
DELF \cite{delf}                      & 74.4      & 71.3      & 78.1         \\
Deeplogo \cite{deeplogo}                  & 78.9      & 72.3      & 73.8       \\   
Logo Proposals \cite{7862773}	 & 69.8	 & 68.2	 & 73.7 \\
One-Shot Network \cite{bhunia2019deep}	 & 71.2	 & 70.4	 & 71.6
\\ 
\hline
\alg~(SIFT) & 73.2      & 79.8      & 78.9         \\
\alg~(DELF) & 80.6      & 80.7      & 83.5         \\
\hline
\end{tabular}}
\end{table}

Table \ref{table:UEFA} shows the mean Average Precision (mAP) of retrieval for each logo target in the UEFA videos, compared to a ground truth which was determined by manually counting the time that each target was visible. \alg~based on DELF achieves more than 80$\%$ mAP over the three year's dataset, which is significantly better than the other models. Both \alg~based on SIFT and DELF improve at least 5$\%$ over the original SIFT and DELF algorithms. Since many regions of the video are irrelevant, models with local features benefit from the small scaled targets. While the logo recognition algorithms show good performance in some cases, they tend to be less consistent and less reliable, likely being influenced by the correlation between the pre-trained dataset and the experimental test dataset.


\begin{table}
\centering
\caption{Comparison of proposed techniques against state-of-the-art algorithms on Stanford I2V dataset in two logo retrieval tasks.}
\label{tab: standford I2V}
\scalebox{0.83}{
\begin{tabular}{c|c|c} 
\hline
\multirow{2}{*}{Model} & \multicolumn{2}{c}{Stanford I2V (mAP)}  \\ 
\cline{2-3}
                       & TV Channel Logo & Social Software Logo  \\ 
\hline
ResNet101~\cite{resnet}         & 79.5            & 30.3                  \\
SIFT~\cite{sift}                   & 83.4            & 38.6                  \\
VLAD \cite{vlad}                   & 82.6            & 41.7                  \\
DELF \cite{delf}                   & 84.5            & 42.3                  \\
Deeplogo \cite{deeplogo}               & 87.1            & 43.1                  \\ 
Logo Proposals \cite{7862773}	 & 84.9	 & 42.5 \\
One-Shot Network \cite{bhunia2019deep}	& 83.5	& 40.7
\\ 
\hline
\alg~(SIFT)              & 87.4            & 41.7                  \\
\alg~(DELF)              & 89.2            & 46.8                  \\
\hline
\end{tabular}}
\end{table}

Table \ref{tab: standford I2V} shows experimental results on the Stanford I2V dataset. The test dataset contains 200 broadcast news videos. Two tasks are tested: 1) Retrieval of three TV channel logos (CNN, CBS and Bloomberg) and 2) Retrieval of social media logos (such as Twitter and Facebook). The \alg~based on DELF achieves the highest mAP for both tasks. \alg~based on SIFT or DELF can achieve improvement of 3$\%$ to 5$\%$ over the original SIFT and DELF algorithms.

\section{conclusion}
This paper introduced a logo image retrieval algorithm based on local features, which consists of three stages: segmentation, matching, and refinement. The algorithm was tested and verified on two benchmark datasets and shows notable performance advantages compared with several state of the art methods. The idea of the \alg~is not limited to SIFT or DELF, and it has potential to improve performance using other local features.


\vfill
\pagebreak

\bibliographystyle{IEEEbib}
\bibliography{strings,refs}

\end{document}